\documentclass[conference,9pt]{IEEEtran}

\IEEEoverridecommandlockouts

\usepackage{graphicx,enumitem, textcase}
\usepackage{amsmath,amssymb,amsthm,bm}
\usepackage{tabularx}
\usepackage{color}

\title{Attention-based multi-task learning for speech-enhancement and speaker-identification in multi-speaker dialogue scenario}
\author{
\IEEEauthorblockN{
Chiang-Jen Peng\IEEEauthorrefmark{1},
Yun-Ju Chan\IEEEauthorrefmark{1},
Cheng Yu\IEEEauthorrefmark{2},
Syu-Siang Wang\IEEEauthorrefmark{2}\IEEEauthorrefmark{3},
Yu Tsao\IEEEauthorrefmark{2} and
Tai-Shih Chi\IEEEauthorrefmark{1}
 }
 \IEEEauthorblockA{\IEEEauthorrefmark{1}Department of Electrical and Computer Engineering, National Chiao Tung University, Hsinchu, Taiwan\\
 Email: tschi@mail.nctu.edu.tw} 
 \IEEEauthorblockA{\IEEEauthorrefmark{2}Research Center for Information Technology Innovation, Academia Sinica, Taipei, Taiwan\\
Email: yu.tsao@citi.sinica.edu.tw, chengyu\_citi@citi.sinica.edu.tw}
 \IEEEauthorblockA{\IEEEauthorrefmark{3}Department of Electrical Engineering, Yuan Ze University, Taoyuan, Taiwan\\
 Email: sypdbhee@saturn.yzu.edu.tw
}
}

\begin{document}

\maketitle

\begin{abstract}
Multi-task learning (MTL) and attention mechanism have been proven to effectively extract robust acoustic features for various speech-related tasks in noisy environments. In this study, we propose an attention-based MTL (ATM) approach that integrates MTL and the attention-weighting mechanism to simultaneously realize a multi-model learning structure that performs speech enhancement (SE) and speaker identification (SI). The proposed ATM system consists of three parts: SE, SI, and attention-Net (AttNet). The SE part is composed of a long-short-term memory (LSTM) model, and a deep neural network (DNN) model is used to develop the SI and AttNet parts. The overall ATM system first extracts the representative features and then enhances the speech signals in LSTM-SE and specifies speaker identity in DNN-SI. The AttNet computes weights based on DNN-SI to prepare better representative features for LSTM-SE. We tested the proposed ATM system on Taiwan Mandarin hearing in noise test sentences. The evaluation results confirmed that the proposed system can effectively enhance speech quality and intelligibility of a given noisy input. Moreover, the accuracy of the SI can also be notably improved by using the proposed ATM system.
\end{abstract}
\section{Introduction}
\label{sec:intro}

Speech signals propagating in an acoustic environment are inevitably distorted by noise. Such distortions may considerably degrade the performance of the target speech-related tasks, such as assistive hearing systems \cite{wang2017deep, lai2016deep} automatic speech recognition \cite{weninger2015speech,li2014overview}, and speaker recognition \cite{kolboek2016speech, shon2019voiceid,a2015automatic}. To address this issue, speech enhancement (SE), which aims to extract clean acoustic signals from noisy inputs, has been widely used. Conventional SE techniques, including signal subspace  \cite{loizou2013speech}, power spectral subtraction \cite{boll1979suppression}, Wiener filtering \cite{lim1979enhancement},
and minimum mean square error based estimations \cite{ephraim1984speech, ephraim1985speech}, perform well in stationary noise environments, where the statistical assumptions on environmental noises and human speech hold adequately \cite{1495469,paliwal2010single,lotter2005speech}. For environments involving non-stationary noises, conventional SE techniques may not provide satisfactory performance. Recently, deep learning (DL)-based SE methods have been widely studied, and notable performance improvements have been observed over conventional techniques. In general, the DL-based SE methods aim to transform the noisy source to a clean target using nonlinear mapping, in which no statistical property assumption of noise and speech acoustic signals is required \cite{xu2014regression,zhao2018two,xu2014dynamic}. In \cite{lu2013speech,yu2020speech}, the authors proposed a deep denoising autoencoder (DDAE) SE system that encodes input noisy signals into a series of frame-level speech codes and then performs a decoding process to retrieve the enhanced 
signals from the system output. Another study in \cite{weninger2015speech} applied a long short-term memory (LSTM) model to integrate the context information to carry out SE for improving speech quality and intelligibility and achieving a low word error rate in an ASR system. In \cite{kim2020t}, the transformer model that utilizes an attention mechanism \cite{vaswani2017attention} to compute attention weights is used to emphasize and fuse related context symbols to obtain clean components.

The SE system can be used as a front-end processor for specific applications by placing it in front of a main speech-signal-processing system. By jointly minimizing the losses from the SE and the main system, the overall system is considered to be optimized in a multi-task learning (MTL) manner \cite{chen2015speech,lee2020multi,shi2019ones}. In such systems, the MTL aims to purify the representations along with the goal to boost the performance of the main task \cite{ruder2017overview,zhang2017survey,crawshaw2020multi}. In \cite{morrone2020audio, hou2018audio}, visual information is treated as the second task to promote the SE capability. Experimental results show that audio and visual cues can be jointly considered to derive more representative acoustic features in a DL-based SE system.

MTL has also been used in speaker recognition, namely speaker identification (SI) and speaker verification (SV), systems \cite{chen2015multi,tang2016multi,pironkov2016speaker}. The recognition accuracy of an SI task is highly dependent on the quality of speaker feature extraction. Therefore, most existing systems aim to compute a decent speaker representation from speech signals. A well-known speaker recognition system is a combination of an $i$-vector with a probabilistic linear discriminant analysis \cite{dehak2010front}. This system has been widely used and yields satisfactory performance in numerous speaker recognition tasks. More recently, $d$-vector \cite{variani2014deep} and $x$-vector \cite{snyder2018x} features extracted by DL models have been proven to provide more abundant speaker information and, therefore, show superior recognition performances to the $i$-vector. 

Inspired by the transformer model structure, this study proposes a novel system, namely the attention-based MTL (ATM), to extract the shared information between SE and SI to attain improved performance for individual tasks. The outputs of the ATM system are enhanced speech and identification results, and the input is noisy speech signals. In addition, an attention-based network (AttNet) is used to integrate both speech and speaker cues between SE and SI models to extract robust features. The ATM consists of three DL-based models: the first LSTM enhances the noisy input and the other two DNNs are used to identify the speaker identity and extract the attention weights. We tested the proposed system on the Taiwan Mandarin hearing in noise test (TMHINT) sentences \cite{huang2005development}. The experimental results show that the proposed ATM can not only enhance the quality and intelligibility of noisy speech but also the improve SI accuracy. 

The remainder of this paper is organized as follows. Section \ref{sec:relwork} reviews the related work, including LSTM-based SE and DNN-based SI. Section \ref{sec:atm} introduces the proposed ATM architecture. Experimental results and analyses are provided in Section \ref{sec:exp}. Finally, Section \ref{sec:concl} presents the conclusions and directions for future research.

\section{RELATED WORKS}\label{sec:relwork}
This section briefly reviews the related works of the LSTM-SE and DNN-SI systems. Considering noisy speech signals are obtained from contaminating clean speech signals with additive noise signals. With short-time Fourier transform (STFT) and several feature processing steps, we can obtain noisy and clean logarithmic power spectra (LPS), $\mathbf{Y}$ and $\mathbf{S}$, respectively, from the noisy and clean speech signals. We assumed that there are $N$ frames in the paired $(\mathbf{Y}$--$\mathbf{S})$. The context feature of noisy LPS by concatenating the adjacent $2M$ feature frames of the target feature vector $\mathbf{Y}[n]$, namely, $\mathbb{Y}[n]=[\mathbf{Y}'[n-M],\cdots,\mathbf{Y}'[n],\cdots,\mathbf{Y}'[n+M]]'$ are used.

\subsection{Speech enhancement}
In this study, the baseline SE system is composed of an $L$-hidden-layer LSTM and a feed-forward layer and is denoted as LSTM-SE. The input-output relationship   ($\mathbf{z}_{\ell+1}[n]$, $\mathbf{z}_\ell[n]$) at the $n$-th frame and the $\ell$-th hidden layer is formulated as
\begin{equation}\label{eq:lstm}
\mathbf{z}_{\ell+1}[n]=LSTM_{\ell}\{\mathbf{z}_{\ell}[n]\}, \:\:\ell=1,2,\cdots,L.
\end{equation}
The input of the first LSTM layer is $\mathbf{Y}$, i.e. $\mathbf{z}_{1}[n]=\mathbf{Y}[n]$, and the output $\mathbf{z}_{L+1}[n]$ is 
\begin{equation}\label{eq:ff}
\hat{\mathbf{S}}[n]=\mathbf{W}\mathbf{z}_{L+1}[n]+\mathbf{b},
\end{equation}
where $\mathbf{W}$ and $\mathbf{b}$ are the weight matrix and bias vector, respectively. In the training stage, the parameters of the LSTM-SE system are updated to minimize the difference between $\hat{\mathbf{S}}[n]$ and $\mathbf{S}[n]$ in terms of the mean square error (MSE). In the testing stage, the output $\hat{\mathbf{S}}$ of the LSTM-SE is combined with the phase from the noisy speech signals to produce the enhanced signals $\hat{\mathbf{s}}$ in the time domain.

\subsection{Speaker identification}\label{sec:si}
The objective of the DNN-SI is to classify an input speech signal $\mathbb{Y}[n]$ at the $n$-th frame to a specific speaker identity. We categorized the non-speech segments as a single virtual speaker. Therefore, the dimension of DNN-SI output is the number of speakers plus one, namely $K+1$. The reference target for the DNN training is a one-hot $(K+1)$-dimensional vector $\mathbf{I}[n]$, where a single non-zero element corresponds to the target speaker identity.

The DNN-SI contains $D$ layers, and the input-output relationship ($\mathbf{z}_{d}[n]$, $\mathbf{z}_{d+1}[n]$) at the $d$-th layer and the $n$-th frame can be formulated as
\begin{equation}\label{eq:dnn}
\mathbf{z}_{d+1}[n]=\sigma_{d}\left(F_{d}(\mathbf{z}_{d}[n])\right),\quad d=1,\cdots,D,
\end{equation}
where $\sigma_{d}(\cdot)$ and $F_{d}(\cdot)$ are the activation and linear transformation functions, respectively. In this study, the softmax function is used as the activation function for the output layer, and the rectified linear units (ReLU) function is used for all hidden layers. Meanwhile, the input and output of DNN is $\mathbf{z}_{1}[n]=\mathbb{Y}[n]$ and $\mathbf{z}_{D+1}[n]=\hat{\mathbf{I}}$, respectively. The categorical cross-entropy loss is used to compute the DNN parameters in Eq. \eqref{eq:dnn}.

\section{The proposed approach}\label{sec:atm}
\begin{figure}[!t]
{\centering\includegraphics[width=\columnwidth]{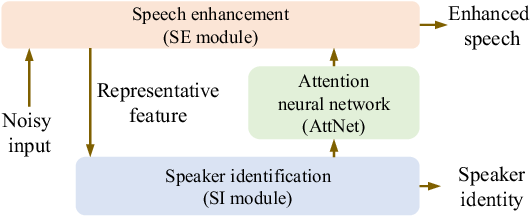}
\caption{The block diagram of the proposed ATM system, in which the input is noisy speech while the outputs are enhanced speech and the recognized speaker identity.}\label{fig:overall}}
\end{figure}

Figure \ref{fig:overall} shows the block diagram of the proposed ATM system. From the figure, the input to the ATM system is a noisy LPS feature, $\mathbf{Y}$, while the outputs have enhanced LPS feature in SE and speaker identity vector in SI. Between SE and SI tasks, an AttNet is employed to reshape the feature size from SI and extract more compact speaker cues for SE. Based on the ways of incorporating attention mechanism, two ATM architectures have been proposed, namely ATM$_{bef}$ and ATM$_{ide}$, which are detailed in the following two sub-sections.

\subsection{The ATM$_{bef}$ system}
\begin{figure}[!t]
\vspace{0.1cm}{\centering\includegraphics[width=0.8\columnwidth]{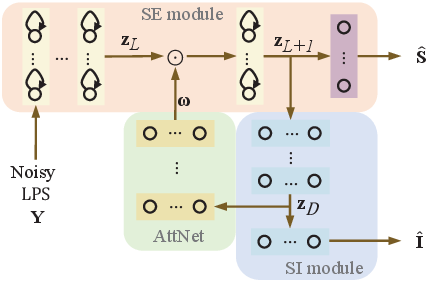}
\caption{The architecture of ATM$_{bef}$. The output of the $(L+1)$-th layer of LSTM-SE is used to compute $\bm{\omega}$, which is then used to weight the representative features at the $L$-th layer.}\label{fig:atmwb}}
\end{figure}

Figure \ref{fig:atmwb} illustrates the block diagram of the ATM$_{bef}$ system. As shown in the figure, the SE model is used to provide the embedded speech code vector, $\mathbf{z}_{L+1}[n]$, from the output of the $L$-th LSTM hidden layer. We then created the context information of speech by concatenating the adjacent vectors of $\mathbf{z}_{L+1}[n]$ to obtain $[\mathbf{z}'_{L+1}[n-M],\cdots;\mathbf{z}'_{L+1}[n],\cdots,\mathbf{z}'_{L+1}[n+M]]'$ to form the input of SI to compute the speaker feature (from the output of the last hidden layer). Then, AttNet, which is a $J$-layer DNN model, takes the speaker feature as the input to compute the weighting vector, $\bm{\omega}$, to weight the LSTM-SE by performing  $\bm{\omega}[n]\odot\mathbf{z}_{L}[n]$, where $\odot$ is an element-wise multiplication operator. Finally, enhanced speech, $\hat{\mathbf{S}}$, and recognized speaker identity, $\hat{\mathbf{I}}$, are obtained. This system is referred to ATM$_{bef}$ because the attention operation is performed before extracting the acoustic feature representation.

To train ATM$_{bef}$, we prepare noisy LPS features as the input and the corresponding speaker-identity vectors and clean LPS features as the two outputs. Then, an iterative training procedure is applied to train SI and SE--AttNet models using the following steps: (1) The categorical cross-entropy loss is used to train the SI model, where the model input and output are the contextual embedding features and the speaker-identity vectors, respectively. (2) The speaker features, $\mathbf{z}_{D}$, using the SI model was then extracted. (3) The training proceeds with $\mathbf{Y}$ and $\mathbf{z}_{D}$ on the input side of SE and AttNet, respectively, to produce an enhanced output that approximates $\mathbf{S}$. Notably, the SE and AttNet models are jointly trained based on the MSE loss.

\subsection{The ATM$_{ide}$ system}
\begin{figure}[!t]
{\centering\includegraphics[width=0.8\columnwidth]{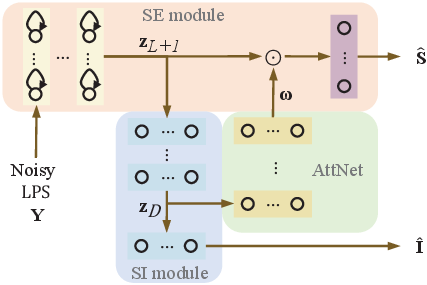}
\caption{The architecture of ATM$_{ide}$. The shared representation of SE and SI and the weighting operation are performed in the same layer of LSTM-SE.}\label{fig:atmaft}}
\end{figure}

Figure \ref{fig:atmaft} shows the block diagram of ATM$_{bef}$. Different from ATM$_{bef}$, ATM$_{ide}$ performs feature extraction and weighting operation in the same layer of LSTM-SE. More specifically, ATM$_{ide}$ performs four steps to obtain $\hat{\mathbf{S}}$ and $\hat{\mathbf{I}}$. First, the acoustic code $\mathbf{z}_{L+1}[n]$ is first computed by passing the noisy LPS $\mathbf{Y}[n]$ to LSTM-SE. Next, the SI model generates $\hat{\mathbf{I}}[n]$ in the output and the speaker code $\mathbf{z}_{D}[n]$, which is served as the input of the AttNet, to obtain the weighting vector, $\bm{\omega}[n]$. Then, the weighting vector, $\bm{\omega}[n]$, is applied to weight the acoustic code (that is, $\bm{\omega}\odot\mathbf{z}_{L+1}[n]$). Finally, a transformation is applied to the weighted acoustic code to generate the enhanced output $\hat{\mathbf{S}}[n]$. For ATM$_{ide}$, the weighting vector $\bm{\omega}$ is extracted to introduce the speaker-dependent characteristics in the acoustic feature and guide the SE to generate the output corresponding to the target speaker. The proposed ATM systems (ATM$_{ide}$ and ATM$_{bef}$) can be considered as multi-model systems because the speaker characteristics are used to guide SE to achieve better performance.

The dynamic weighted loss function has been proposed to address the scaling issue of classification and regression tasks \cite{kendall2018multi,sener2018multi}. The loss is formulated in Eq. \eqref{eq:dymweiloss} with two additional trainable parameters, $\alpha$ and $\beta$.
\begin{equation}\label{eq:dymweiloss}
L(\Theta,\alpha,\beta)=\frac{1}{2\alpha^2}L_1(\Theta) + \frac{1}{\beta^2}L_2(\Theta) + log\alpha + log\beta,
\end{equation}
where $L_1$ and $L_2$ are the MSE and the categorical cross-entropy loss, respectively; $\Theta$ represents all the parameters in ATM$_{ide}$. 

\section{Experiments and Analyses}\label{sec:exp}
In the following subsections, we first introduce the experimental setup and then provide the experimental results along with a discussion about our findings.

\subsection{Experimental setup}
We evaluated the proposed ATM system on TMHINT sentences. The training and testing utterances were recorded by eight speakers at a 16 kHz sampling rate in a noise-free meeting room. A total of 1,560 clean utterances were pronounced by three males and three females ($K=6$ in Section \ref{sec:si}), each providing 260 utterances, for the training set. From these clean utterances, we randomly concatenated three utterances to simulate the dialogue scenario and subsequently generated 520 clean training utterances, where each utterance contained exactly three different speakers. Noisy utterances were generated by artificially adding 100 different types of noises \cite{100noise} at six signal-to-noise ratio (SNR) levels ($15$, $10$, $5$, $0$, $-5$, and $-10$ dB) to the prepared 520 clean training utterances, and thus generating 312,000 ($=520\times 100\times 6$) noisy--clean training pairs. Among them, we randomly selected 500 noisy--clean pairs to form the validation set. Meanwhile, two different testing configurations were prepared for the SE and SI tasks. For SE, the testing set contained additional male and female speech. We randomly concatenated one utterance of the male speaker and one utterance of the female speaker and generated 60 clean utterances for testing. Noisy testing utterances were then prepared by deteriorating these clean utterances with four additive noises (``engine'', ``pink'', ``street'', and ``white'') at three SNRs ($5$, $0$, and $-5$ dBs). Accordingly, we prepared 720 ($=60\times 4\times 3$) noisy testing utterances. In contrast to the SE testing set, the utterances used to evaluate the SI part were from the same speakers in the training set. We prepared 120 clean dialogue utterances for testing, with each utterance containing segments from three different speakers. Then, we added four additive noises (``engine'', ``pink'', ``street'', and ``white'') at three SNRs ($5$, $0$, and $-5$ dBs) to these clean testing utterances to form the noisy utterances. Accordingly, we prepared 1440 noisy utterances for testing the SI performance. Please note that we did not consider overlapped speech in this study, and thus there were no overlapped segments in the utterances for all the training and testing sets.

To apply the STFT, we used a window with frame size and shift of 32 ms and 16 ms, respectively. Then, a 257-dimensional LPS vector was obtained. The context feature was created by $M=5$ thus with the dimension of 2,827$=257\times(2\times 5+1)$. Accordingly, the input- and output-layer sizes of SE were both 257, and those of SI were 2,827 and 7 (i.e., $K+1=6+1$), respectively. For the overall ATM system, the input size were 257 and the output sizes was 257 for SE and 7 for SI. The detailed network configuration is as follows:
\begin{itemize}
	\item The SE model consisted of two LSTM layers ($L=2$) with 300 cells in each layer, followed by a 257-node feed-forward layer.
	\item The SI model comprised four hidden layers ($D=4$) in the order of 1024, 1024, 256, and 7 nodes.
	\item The AttNet comprised two hidden layers ($J=2$) with each layer having 300 nodes.  
\end{itemize}

In this study, we applied three metrics to evaluate the proposed system: perceptual evaluation of speech quality (PESQ) \cite{rix2001perceptual}, short-time objective intelligibility (STOI) \cite{taal2011algorithm}, and segmental SNR index (SSNRI) \cite{chen2008fundamentals}. The score ranges of PESQ and STOI are $[-0.5, 4.5]$ and $[0, 1]$, respectively. Higher PESQ and STOI scores indicate better speech quality and intelligibility. Meanwhile, a higher SSNRI score indicates better signal-level SE performance. 

\begin{table}[b]
\caption{Averaged PESQ, STOI and SSNRI scores of Noisy, LSTM-SE, MTL, ATM$_{\textit{bef}}$, and ATM$_{\textit{ide}}$.}\label{tab:avgse}
\begin{center}
\begin{tabular}{c||c|c|c|c|c}
\hline
& \textbf{Noisy} & \textbf{LSTM-SE} & \textbf{MTL} & \textbf{ATM$_{\textit{bef}}$} & \textbf{ATM$_{\textit{ide}}$} \\
\hline
\textbf{PESQ} & 1.25 & 1.86 & 1.86 & 1.94 & \textbf{1.98} \\ 
\hline
\textbf{STOI} & 0.72 & 0.73 & 0.74 & 0.74 & \textbf{0.75} \\
\hline
\textbf{SSNRI}& -- & 7.39 & 7.61 & 7.57 & \textbf{8.05}\\
\hline
\hline
\end{tabular}
\end{center}
\end{table}
\begin{table}[b]
\caption{The averaged scores of PESQ with respect to four different noise environments over all SNR levels, achieved by Noisy, LSTM-SE, MTL, ATM$_{\textit{bef}}$, and ATM$_{\textit{ide}}$ systems.}\label{tab:pesq}
\begin{center} 
\begin{tabular}{c||c|c|c|c|c}
\hline
& \textbf{Noisy} & \textbf{LSTM-SE} & \textbf{MTL} & \textbf{ATM$_{\textit{bef}}$} & \textbf{ATM$_{\textit{ide}}$} \\
\hline
WHITE & 1.25 & 2.01 & 2.00 & 2.08 & \textbf{2.13} \\
\hline
PINK & 1.28 & 1.88 & 1.88 & 1.96 & \textbf{2.02} \\
\hline
STREET & 1.32 & 1.84 & 1.83 & 1.89 & \textbf{1.92} \\
\hline
ENGINE & 1.16 & 1.72 & 1.71 & 1.81 & \textbf{1.84} \\
\hline
\hline
\end{tabular} 
\end{center}
\end{table}
\begin{table}[!pb]
\caption{The averaged scores of STOI with respect to different noise environments over all SNR levels, achieved by Noisy, LSTM-SE, MTL, ATM$_{\textit{bef}}$, and ATM$_{\textit{ide}}$.}\label{tab:stoi}
\begin{center} 
\begin{tabular}{c||c|c|c|c|c}
\hline
& \textbf{Noisy} & \textbf{LSTM-SE} & \textbf{MTL} & \textbf{ATM$_{\textit{bef}}$} & \textbf{ATM$_{\textit{ide}}$} \\
\hline
WHITE & 0.75 & 0.75 & 0.75 & 0.76 & \textbf{0.77} \\
\hline
PINK & 0.72 & 0.72 & 0.73 & 0.73 & \textbf{0.74} \\
\hline
STREET & 0.72 & 0.74 & 0.75 & 0.75 & \textbf{0.76} \\
\hline
ENGINE & 0.69 & 0.70 & 0.71 & 0.71 & \textbf{0.73} \\
\hline
\hline
\end{tabular} 
\end{center}
\end{table}
\subsection{Experimental results}
In this subsection, we split the evaluation results into two parts. We first report the SE evaluation results and then the SI performance.
\subsubsection{SE results}
Table \ref{tab:avgse} lists the averaged PESQ, STOI, and SSNRI results with respect to all testing utterances of the noisy baseline (denoted as ``Noisy'') and the enhanced speech obtained by conventional LSTM-SE, ATM$_{\textit{bef}}$, and ATM$_{\textit{ide}}$. In addition, the results of MTL, which is composed of only SE and SI models (without AttNet) in Fig. \ref{fig:overall}, are also listed for comparison. From the table, most evaluation metrics using the MTL criterion, that is, MTL, ATM$_{\textit{bef}}$, and ATM$_{\textit{ide}}$, show better results than those provided by LSTM-SE, except the PESQ score of MTL. The results confirm the effectiveness of MTL-based models in improving the speech quality, intelligibility, and background noise reductions. In addition, both ATM$_{\textit{bef}}$ and ATM$_{\textit{ide}}$ provide better results than MTL for all evaluation metrics. The results confirm that the MLT-based SE system can be further improved by applying the attention-weighting mechanism. In addition, ATM$_{\textit{ide}}$ yields scores superior to ATM$_{\textit{bef}}$ implying that a suitable attention mechanism further promotes the system capability.

To further analyze the benefits of the proposed systems, we report the detailed PESQ and STOI scores of Table \ref{tab:avgse} in Tables \ref{tab:pesq} and \ref{tab:stoi}, respectively. We compared the performance of Noisy, LSTM, MTL, ATM$_{\textit{bef}}$, and ATM$_{\textit{ide}}$ with respect to four testing noise environments over all SNR levels. From both tables, we observe that all DL-based SE approaches provide better PESQ and STOI scores on all evaluated conditions than the noisy baseline, while ATM$_{\textit{ide}}$ performs the best. The results verify the capability of the proposed ATM approach to extract robust features for SE, thus further improving speech quality and intelligibility. 

\begin{figure}[t]
\centering
\includegraphics[width=0.85\columnwidth]{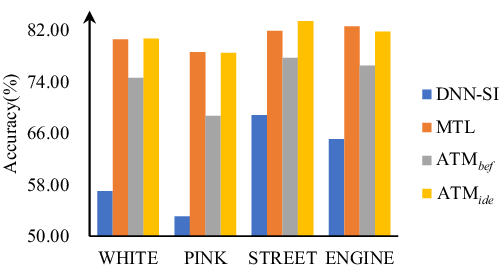}
\caption{The frame-wise SI accuracy of DNN-SI, MTL, ATM$_{\textit{bef}}$, and ATM$_{\textit{ide}}$ in four testing noise environments.}
\label{fig:acc}
\end{figure}
\begin{figure}[!pt]
\begin{center}
\vspace{0.25cm}\includegraphics[width=\columnwidth]{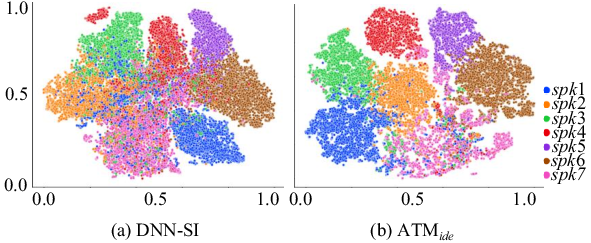}
\end{center}
\vspace{-0.2cm}\caption{The distributions of (a) DNN-SI and (b) ATM$_{\textit{ide}}$ extracted speaker features with t-SNE analyses.}
\label{fig:t-sne}
\end{figure}

\subsubsection{SI results}
Figure \ref{fig:acc} illustrates the frame-wise SI accuracy of the DNN-SI baseline, MTL, ATM$_{\textit{bef}}$, and ATM$_{\textit{ide}}$. The evaluations were conducted on testing utterances involving ``engine'', ``pink'', ``street'', and ``white'' noise backgrounds, among which ``street'' is considered to be the most complex noise type. From the figure, it can be observed that MTL-based approaches (MTL, ATM$_{\textit{bef}}$, and ATM$_{\textit{ide}}$) provide higher SI accuracies than those achieved by conventional DNN-SI. In addition, ATM$_{\textit{ide}}$ shows the highest recognition accuracy in the street background, and competes with MTL in other noise environments. The results demonstrate that the MTL architecture can effectively enhance the SI performance and can be further improved by incorporating the attention-weighting mechanism. 

Next, we analyze the speaker features using DNN-SI and ATM$_{\textit{ide}}$ based on t-SNE analyses \cite{maaten2008visualizing}. The t-SNE analysis is a widely used technique that provides visualized feature clusters from high-dimensional spaces. In this study, seven speakers were involved in the training set (including one non-speech virtual speaker). The analysis was carried out by first placing all SI-testing noisy utterances on the input of DNN-SI or ATM$_{\textit{ide}}$ to derive the associated speaker features. Then, these high-dimensional DNN-SI- and ATM$_{\textit{ide}}$-extracted speaker features were processed by t-SNE to yield two-dimensional representations. Fig. \ref{fig:t-sne} illustrates the distributions of these dimension-reduced (a) DNN-SI and (b) ATM$_{\textit{ide}}$ features with associated speaker identities. In the figure, it can be observed that the ATM$_{\textit{ide}}$ system provides a larger inter-class distance and a clearer class boundary than those of the DNN-SI baseline. The results show that a combination of MTL and AttNet techniques can extract more representative features for the SI task. 

\section{conclusion}\label{sec:concl}
In this study, we proposed a novel ATM approach that integrates the MTL and the attention-weighting mechanism to carry out SE and SI tasks simultaneously. The overall ATM system is composed of SE, SI, and AttNet modules, and is able to extract representative and robust acoustic features in a noisy environment. Experimental results on the simulated dialog conditions confirm that the proposed ATM can significantly reduce the noise components from the noisy speech, thereby improving speech quality and intelligibility for the SE task. Meanwhile, a suitable attention mechanism performed in ATM could further improve the enhancement performance. On the other hand, the recognition accuracy of the SI system can be further improved through the proposed ATM approach. In the future, we plan to test the ATM system with other languages. We will also explore ATM by using other types of SE and SI models. Finally, the we will test the proposed ATM architecture on speaker-diarization and speech-source separation tasks.

\vfill\pagebreak

\bibliographystyle{IEEEbib}
\bibliography{refs}

\end{document}